\documentstyle[aps,preprint]{revtex}
\begin{document}
\title{Shape transition in some rare-earth nuclei in 
relativistic mean field theory}
\author{B. K. Agrawal$^{1}$\footnote{E-mail:bijay@tnp.saha.ernet.in}, 
Tapas Sil$^{2}$, S. K. Samaddar$^1$ and J. N. De$^2$}

\address{$^{(1)}$ Saha Institute of Nuclear Physics,
1/AF, Bidhannagar, Calcutta - 700064, India}

\address{$^{(2)}$ Variable Energy Cyclotron Centre,
1/AF, Bidhannagar, Calcutta - 700064, India}

\date{\today}

\maketitle

\begin{abstract}
A systematic study of the temperature dependence of the shapes
and pairing gaps of some isotopes in the rare-earth region is
made in the relativistic Hartree-BCS theory. 
Thermal response  to these nuclei is always  found to  lead to 
a phase transition from the superfluid to the normal phase 
at a temperature  $T_{\Delta}\sim 0.4 - 0.8$ MeV and a shape
transition from prolate to spherical shapes at $T_c\sim 1.0 - 2.5$ MeV. These
shape transition temperatures are appreciably higher than the corresponding
ones calculated in the non-relativistic framework with the  pairing plus 
quadrupole interaction. Study of nuclei with  continued addition
of  neutron pairs  for a given isotope shows that
with increased ground state deformation, the transition to the
spherical shape is delayed in temperature. 
A strong linear correlation between $T_{\Delta}$ and the ground
state pairing gap $\Delta^0$ is observed; a well-marked linear
correlation between $T_c$ and the ground state quadrupole 
defromation $\beta_2^{0}$ is also seen. 
The  thermal evolution of 
the hexadecapole deformation is further presented in the paper.
\vskip .5 cm

\end{abstract}

\newpage
\section{Introduction}
Heating can have a  profound effect on nuclear shapes, 
causing a variety of  shape transitions. Experimentally, such  responses
to the thermal excitations have been studied from the shapes of the giant dipole
resonances(GDR) built on excited states \cite{sno,gaa,nan}. Theoretically,
they have been studied earlier in a finite temperature
non-relativistic microscopic Hartree-Fock \cite{bra1,que} and 
Hartree-Fock Bogoliubov (HFB) framework \cite{goo0,goo1,goo2,egi1} with
a pairing plus quadrupole (P+Q) interaction. For the nuclei studied,
it has been found that  while the superfluid  nuclear phase  has transition to 
the normal phase typically at T$\sim 0.5 $ MeV, the deformed shapes  have 
transition  to the spherical
ones  at a higher temperature, mostly between T$\sim 1.0 $ to 1.8 MeV
for rare-earth nuclei. These calculations,  however, have some limitations;
they employ a simplistic model Hamiltonian in a limited model space,  an 
inert core is assumed, moreover, the Coulomb interaction has not been taken
into account  realistically. The understanding  of the universal pattern
of the mean field shape evolution with temperature has 
also been tried in a macroscopic 
approach \cite{lev,alha} commonly referred to as the Landau theory of phase
transition. A quantitative estimate of the persistence of the
ground state deformation  \cite{kel} with 
temperature is however seen  to be missing in some cases.

Recently,  we have undertaken a study  \cite{agr} of the 
thermal evolution of nuclear 
properties,  particularly the phase transition in the  nuclear shape 
and  the superfluidity in the relativistic mean field (RMF) theory.  
The pairing effects have been included in the BCS approximation.
The RMF theory  \cite{ser,gam,rin} has proved to be an extremely powerful tool  
in explaining  the  gross properties of nuclei over the entire periodic table. 
In contrast to the non-relativistic models, this theory employs a single
set of parameters to explain  all these properties. Moreover, in 
such calculations  , the model space used is sufficiently large and all
the nucleons are treated on equal footing. The calculations reported in 
Ref. \cite{agr} are performed  for only two  rare-earth 
nuclei, namely, $^{166}Er$ and $^{170}Er$. It is
found that  the phase transition  for the nuclear shape 
from the prolate to the spherical occurs
at a temperature significantly higher than that obtained in the (P+Q) model and
that the transition is  relatively smooth. A very recent calculation by
Egido et al \cite{egi} in a nonrelativistic approach but with the realistic
Gogny force reveals that the characteristics of  
the nuclear shape transition are very similar to those 
obtained  in the RMF theory.
We have  therefore undertaken a  more systematic 
study  of the shape transition for the
rare-earth nuclei in the  relativistic mean field theory in the
present paper.  For this purpose, we have considered  various
even-even isotopes  of $Sm$, $Gd$ and  $Dy$. A more quantitative study,
particularly of the shape transitions, calls for the inclusion of
thermal fluctuations \cite{agr,alha1,ros}. However, 
this is too computer intensive and
are not included in the present paper.

The organization of the paper is as follows:  we discuss the
theoretical framework briefly in section II. The results and discussions are
presented in section III and the  concluding remarks are given in  section IV.

\section{Formalism}
We employ the nonlinear $\sigma-\omega-\rho$ version of the RMF theory
\cite{gam}.
The Lagrangian density  for the nucleon-meson many body system is taken as
\begin{eqnarray}
\label{lag}
{\cal L}&=& \bar\Psi_i\left ( i\gamma^\mu \partial_\mu - M\right )\Psi_i
+ \frac{1}{2} \partial^\mu\sigma\partial_\mu\sigma - U(\sigma)
- g_\sigma \bar\Psi_i \sigma\Psi_i\nonumber\\
&& - \frac{1}{4}\Omega^{\mu\nu}\Omega_{\mu\nu}
+\frac{1}{2}m_\omega^2\omega^\mu \omega_\mu - g_\omega \bar\Psi_i \gamma^\mu
\omega_\mu\Psi_i 
-\frac{1}{4}\vec{R}^{\mu\nu} \vec{R}_{\mu\nu} + \frac{1}{2} m_\rho^2 \vec{\rho}^\mu\vec{\rho}_\mu\nonumber\\
&&- g_\rho \bar\Psi_i \gamma^\mu\vec{\rho}_\mu\vec{\tau}\Psi_i
-\frac{1}{4}F^{\mu\nu}F_{\mu\nu} - e\bar\Psi_i \gamma^\mu \frac{(1-\tau_3)}{2} A_\mu\Psi_i
.
\label{Lag}
\end{eqnarray}

The arrows indicate  isovector quantities. The mesons
included  in the  description
are the $\sigma$, $\omega$ and $\rho$ mesons. 
For an appropriate description of the nuclear surfaces
\cite{bog},  a non-linear scalar self-interaction  
term $U(\sigma)$  of the $\sigma$ meson
is included in the Lagrangian
\begin{equation}
U(\sigma) = \frac{1}{2}m_\sigma^2 \sigma^2 + \frac{1}{3}g_2 \sigma^3 + 
\frac{1}{4}g_3\sigma^4
.
\end{equation}

The meson masses are given by  $m_\sigma$, $m_\omega$ and $m_\rho$, the nucleon mass is
$M$ and $g_\sigma$, $g_\omega$, $g_\rho$ and $e^2/4\pi=1/137$ are the 
coupling constants for the mesons  and the photon. 
The field tensors for the vector mesons 
$\omega$ and $\rho$ are given by $\Omega^{\mu\nu}$  and  
$\vec R^{\mu\nu}$, for the
electromagnetic field, it is  $F^{\mu\nu}$.  
Recourse to variational principle followed by
the mean field approximation treating the fields as $c-$ numbers results in the
Dirac equation for the nucleon  and the Klein-Gordon type equations  
for the mesons and the 
photon. For the static case, along with the  time-reversal  invariance
and charge conservation, the equations get simplified. 
The resulting equations, known as RMF equations, have the following form.
The Dirac equation  for the nucleon is  
\begin{equation}
\label{dir}
\{-i{\bf\alpha}\cdot{\bf \nabla} + V({\bf r}) +\beta\left [M+S({\bf r})
\right ] \} \Psi_i = \epsilon_i\Psi_i,
\end{equation}
where $V({\bf r})$ represents  the {\it vector } potential 
\begin{equation}
V({\bf r}) = g_\omega \omega_0({\bf r})  + g_\rho \tau_3{\bf \rho}_0({\bf r})
+ e\frac{(1 - \tau_3)}{2} A_0({\bf r}),
\end{equation}
and $S({\bf r})$ is the {\it scalar} potential 
\begin{equation}
S({\bf r}) = g_\sigma\sigma({\bf r}),
\end{equation}
which contributes  to the effective mass as 
\begin{equation}
M^*({\bf r}) = M + S({\bf r})
.
\end{equation}

The Klein-Gordon equations for the mesons and the electromagnetic fields  with the
nucleon densities as sources  are 
\begin{eqnarray}
\label{sig}
\left \{ -\Delta + m_\sigma^2\right \} \sigma({\bf r}) & = & -g_\sigma\rho_s({\bf r})
-g_2\sigma^2({\bf r}) - g_3\sigma^3({\bf r}),\\
\label{ome}
\left \{ -\Delta + m_\omega^2\right \} \omega_0({\bf r}) 
& = & g_\omega \rho_v({\bf r}),\\
\label{rho}
\left \{-\Delta +  m_\rho^2\right \} \rho_0({\bf r})  & = & g_\rho\rho_3({\bf r}),\\
\label{pho}
-\Delta A_0({\bf r}) & = & e\rho_c({\bf r}).
\end{eqnarray}
The corresponding densities are  
\begin{eqnarray}
\rho_s & = & \sum_i n_i\bar\Psi_i \Psi_i,\nonumber\\
\rho_v & = & \sum_i n_i\Psi^\dagger_i \Psi_i,\nonumber\\
\rho_3 & = & \sum_i n_i\Psi^\dagger_i\tau_3 \Psi_i,\nonumber\\
\rho_c & = & \sum_i n_i\Psi^\dagger_i\frac{(1-\tau_3)}{2} \Psi_i.
\end{eqnarray}
Here the sums are taken over the particle states only, i.e., 
the negative-energy states are neglected. 
The partial occupancy ($n_i$)
at finite temperature in the BCS approximation  is 
\begin{equation}
n_i = \frac{1}{2}\left [1-\frac{\epsilon_i-\lambda}{\tilde\epsilon_i}
\left ( 1 - 2f(\tilde\epsilon_i,T)\right )\right ],
\end{equation}
with 
$f(\tilde\epsilon_i,T) = 1/(1 + e^{\tilde\epsilon_i/T})$;
$\tilde\epsilon_i = \sqrt{(\epsilon_i - \lambda)^2 + \Delta^2}$
is the quasiparticle energy 
 where $\epsilon_i$ 
is the single-particle energy for the state $i$.  The chemical potential
$\lambda$
for protons (neutrons) is obtained from the requirement
\begin{equation}
\sum_i n_i = Z \> (\> N)
\end{equation}
The sum is taken over proton (neutron) states.
The gap parameter $\Delta$ is obtained by minimising the free energy
\begin{equation}
F = E - TS,
\end{equation}
where
\begin{equation}
\label{ene}
E(T) = \sum_i \epsilon_i n_i + E_\sigma + E_{\sigma NL}+E_\omega + E_\rho 
+ E_C +E_{pair} + E_{c.m.} - AM,
\end{equation}
and 
\begin{equation}
S = -\sum_i \left [ f_i lnf_i + (1-f_i) ln(1-f_i)\right ],
\end{equation}
with
\begin{eqnarray}
\label{esi}
E_\sigma& = &-\frac{1}{2}g_\sigma \int d^3r \rho_s({\bf r}) 
\sigma({\bf r}),\\
\label{esinl}
E_{\sigma NL}& = &-\frac{1}{2} \int d^3r
\left\{\frac{1}{3}g_2 \sigma^3({\bf r})+\frac{1}
{2}g_3\sigma^4({\bf r})\right\},\\
\label{emo}
E_\omega& = &-\frac{1}{2}g_\omega\int d^3r \rho_v({\bf r})\omega^0({\bf r}),\\
\label{erh}
E_\rho& = &-\frac{1}{2}g_\rho\int d^3 r \rho_3({\bf r}) \rho^0({\bf r}),\\
\label{ecou}
E_C& = &-\frac{e^2}{8\pi}\int d^3r \rho_C({\bf r}) A^0({\bf r}),\\
\label{epair}
E_{pair}& = &-\frac{\Delta^2}{G},\\
\label{ecm}
E_{c.m.}& = &-\frac{3}{4}\hbar\omega_0 = -\frac{3}{4}41 A^{-1/3}.
\end{eqnarray}
Here $G$ and $A$ are the pairing strength and the mass number respectively.
The single-particle energies  and the fields appearing in eqs. (\ref{ene}) 
- (\ref{ecou}) are obtained  from the self-consistent solution of 
eqs. (\ref{dir}) - (\ref{pho}). 
The temperature dependent occupancies of the fermions induce temperature
dependence in the bosonic fields through the source terms as seen from
eqs. (\ref{sig})-(\ref{pho}).

We generate these  self-consistent solutions  using the basis expansion
method \cite{gam,rin1} ; this yields the quadrupole deformation 
$\beta_2$, the hexadecapole deformation $\beta_4$ and the proton 
and neutron pairing gaps $\Delta_p$ and $\Delta_n$  as a function
of temperature.

\section{Results and discussions}

We have chosen  even-even isotopes  of the nuclei $Sm$,
$Gd$ and $Dy$ for the study of the pairing  and shape transitions. The NL3
parameter set  is chosen for  the  values of the  coupling constants  and the
masses of the mesons and the nucleons. This parameter set reproduces 
best the ground state
as well as the compression properties of finite nuclei simultaneously
\cite{lal1};  
however, it has already been reported in Ref. \cite{agr}
that the results for shape transition are not that sensitive to the
choice of the  parameter set. The pairing gaps  
$\Delta_n^{0}$ and $\Delta_p^{0}$ for
neutrons and protons for a nucleus  in the ground 
state  are determined from the 
experimental odd-even mass differences \cite{boh}. 
The single-particle states are calculated using spherical oscillator basis 
with twelve shells.
The values of the chemical
potential and  the pairing gap at a given temperature are determined
using  all the single  particle states upto  $2\hbar\omega_0$
(the model space) above the Fermi surface without assuming any core.  

At finite temperature, because of the partial occupancies 
of nucleons above the Fermi surface, it is in principle
necessary to have a larger basis and an extended model space.
It is further necessary to take effects due to continuum 
into account \cite{bon}.
In order to check the convergence of the calculations, we have enlarged 
the basis space from twelve shells to twenty shells and have 
extended the model space to include  single-particle states upto 
$3\hbar\omega_0$ above the Fermi surface. For this
extended model space, the pairing strength $G$ is
adjusted to reproduce the ground state pairing gap. 
The changes in the values of the observables are found to be insignificant 
due to this extension of the basis and  model space even at the
highest temperature of our interest ($T\sim 3.0 $ MeV). To estimate 
the importance of  the continuum corrections on the observables we report
here, we calculated the occupancy $n^{(+)}$ of the single particle states with 
positive energy.  For $T < 1$ MeV, practically there is no particle
in the positive energy states ($n^{(+)} = 0$) and at the
highest temperature of interest studied here 
($T = 2.7$ MeV), $n^{(+)}/A = 0.011$ which is very small.
It is therefore expected that  continuum  corrections may not play
an important role in the temperature range we study. 
Calculations of nuclear level density 
in earlier studies \cite{agr,agr2}, have shown that the 
continuum corrections are not important for
$T $ upto $\sim 3$ MeV. The continuum effects may grow stronger
for $T > 3$ MeV, however, this is beyond the shape
transition temperatures and so we have not taken this into account.

The temperature evolution  of  the quadrupole deformation $\beta_2$
with neutron number $N=86$  and 88  for the systems $Sm$, $Gd$ and $Dy$  are
displayed in Fig. 1. It is well known that  addition of 
nucleons beyond the closed
shell  gives nuclei   progressive prolate deformation upto
around the middle of the next shell closure. 
This is  reflected in the figure for  all the
isotopes and isotones.  It is also seen that
the critical temperature increases  
with addition of  nucleons  for these systems. It is not
immediately apparent whether there is a  close correlation between the
ground state deformation $\beta_2^{0}$ and the critical 
temperature $T_c$ for
shape transition;  we come back to this issue
later. In the top panel of this figure, particularly for
$^{148}Sm$, it is seen that the deformation increases a little
with  temperature before finally falling to zero. This is due to  the delicate
balance between  the temperature dependence of the pairing force and the
nuclear interaction as  derived from the RMF theory.  
The dramatic build-up of
a deformation  in  a temperature window for this 
nucleus  as seen earlier \cite{goo0,goo1} in the non-relativistic 
framework  is  absent in
our calculations.  

The temperature dependence  of the pairing gaps  
$\Delta_p$ and $\Delta_n$  for protons
and neutrons  for the two isotopes each of $Sm$, $Gd$  and $Dy$  are
shown in Figs. 2 and 3.  The pairing gaps  decrease monotonically 
with temperature,  vanishing at  $T\sim 0.6 - 0.7$ MeV for neutrons
and at  $\sim 0.65 - 0.85$ MeV for protons.
The sudden collapse of the pairing gap  and the nuclear deformation 
at some specified temperatures  signifies  phase transitions;  these
correspond to  transition from the superfluid nuclear phase
to the normal phase and a transition from the deformed shape
to  the spherical shape, respectively.  These transitions show
up as bumps  (displayed in Fig. 4  for  $^{148}Sm$ and $^{150}Sm$) 
in the temperature  evolution  of  the heat capacity defined as  
\begin{equation}
C(T) = \frac{\partial E^*}{\partial T}
\end{equation}
where $E^*$  is the excitation energy of the nucleus in
question. At  a temperature $T_\Delta\sim 0.6$ MeV, the  twin peaks 
are seen for both the nuclei referring to the 
dissolution  of the  neutron and proton pairing gaps.  These
are the characteristic signatures  of 
second order phase transition from the superfluid to
the normal phase.  A somewhat more  prominent bump 
is seen  at a temperature  $T_c\sim 1.15$ MeV for 
$^{148}Sm$  (upper panel). This corresponds 
to the  nuclear shape transition.  Addition of two neutrons  (lower panel
for $^{150}Sm$)  shifts the shape transition temperature to 
$T_c\sim 1.6$ MeV.  This is possibly 
due to the larger ground state deformation  of the $^{150}Sm$ nucleus.

From the study of the hot $^{148}Sm$
and $^{150}Sm$ nuclei , it was conjectured earlier 
\cite{goo1} that  addition of two neutrons
might increase the critical temperature for 
deformation collapse. To test this conjecture in
detail, we have calculated the ground state  
quadrupole deformation and shape transition temperatures for
a host of even-even $Sm$ isotopes. The results 
are displayed in Fig. 5. In the vicinity of
the closed shell  ($N=82$),  the ground state quadrupole 
deformation $\beta_2^{0}$ increases  fast with the 
addition of two neutrons as seen from the top panel of Fig. 5.  
As the neutron number approaches 
the mid shell, the deformation levels off, and  
then,  as is well known,  switches over to the oblate shape
\cite{lal2} (not shown in the figure).  The critical 
temperature ($T_c$) for the shape transition  also
increases  with  neutron pair  addition as seen 
in the bottom panel of Fig. 5. The functional behavior of $\beta_2^{0}$ and
$T_c$  with neutron number $N$ are found to be very similar . Indeed there
is a  strong correlation between $\beta_2^{0}$ and $T_c$ as displayed in
Fig. 6.  The filled circles refer to  the  results from the  present
calculation;  they can  be fairly well fitted 
with a straight line  
\begin{equation}
T_c = 7.75 \beta_2^{0}.
\end{equation}
The fit is obtained using results from the twelve different isotopes
of $Sm$, $Gd$, $Dy$ and $Er$ nuclei.
The results calculated \cite{goo1,goo2,goo3} in the
(P+Q) model are also presented in the figure (open squares) 
for a comparison with those
obtained from  the  RMF theory. The $T_c - \beta_2^{0}$ 
correlation is  then also found  to be
approximately linear with a  smaller slope.
The shape transition temperatures obtained in the
present calculations are somewhat higher compared to those obtained
in the $(P+Q)$ model; one may be inclined to attribute
this difference to the higher effective mass  in the $(P+Q)$ model. 
However, calculations with different sets of field  parameters in the
RMF theory with considerably different values of the
effective masses  yield conflicting results.
We have done calculations for $^{150}Sm$ with the parameter sets HS and NL2
which yield very different $M^*/M$ \cite{gam} (0.54 and 0.67), but the
$T_c$ comes out to be 1.45 and 1.75 MeV, respectively, contrary to the
simple-minded expectations. 
Therefore no simple explanation for the higher values of $T_c$ in the
RMF theory is obvious.

How does the pairing transition temperature $T_\Delta$ depend on 
the pairing gap at $T = 0$?  Intuitively 
one would expect  the collapse of nuclear superfluidity at a larger temperature 
if the ground state pairing  gap $\Delta^{0}$ is larger. 
To study it quantitatively,  we have
done calculations for the different isotopes of $Sm$. The results are
shown in Fig. 7. In its upper panel, we display the experimental neutron and
proton ground state pairing gaps as a function of neutron number  for the
$Sm$ isotopes; in the lower panel, the pairing transition temperatures 
$T_\Delta$ are displayed.  An extremely strong correlation 
between $T_\Delta$ and $\Delta^{0}$
for both neutrons and protons is seen;  this is
manifest in the linear relationship 
\begin{equation}
T_{\Delta_p} = 0.56 \Delta_p^{0},\qquad \qquad
T_{\Delta_n} = 0.60 \Delta_n^{0},
\end{equation}
for both neutrons and protons which is also shown in Fig.8.
The points in the
figure include, in addition to  $Sm$,
results from $Gd$, $Dy$ and $Er$ isotopes. 
The relation between $T_\Delta$ and $\Delta^{0}$ is in very close agreement
to that obtained in the nonrelativistic (P+Q) model \cite{goo1}.

The hexadecapole deformation $\beta_4$, if any, also collapses at the same shape
transition temperature $T_c$ as the  quadrupole deformation.  
In Fig. 9, we plot the  hexadecapole moment
(a measure of $\beta_4$) as a  function of temperature for  $^{148}Sm$ and
$^{150}Sm$. Addition of neutrons gives  larger ground state 
$\beta_4^{0}$. The deformation increases smoothly at
low temperature  upto  $T\sim 0.7$ MeV and  then collapses to
zero at $T_c$.  The initial enhancement of this deformation is related to the
weakening  of the pairing correlations with temperature. Such an
enhancement is also seen in a  finite temperature 
Hartree-Fock Bogoliubov (FTHFB) 
calculation  using the finite range density-dependent Gogny force \cite{egi}. 

\section{Conclusions}

The relativistic mean field theory  has been applied to 
understand  properties of  some rare-earth  even-even nuclei at finite
temperature. Pairing effects  have been included through the 
BCS approximation. Focus is made mainly on the temperature-induced 
transition from
the nuclear superfluid phase to the normal phase  and  also  on the
shape transition from a deformed shape  to a spherical one. 
To find out the  systematics  of the dependence  
of the pairing and  shape transition temperatures on
the values of  the ground   state pairing gap  and the  ground state
deformation, we have done calculations for several isotopes 
of  different rare-earth nuclei.  We find that  there is a 
linear correlation between the above mentioned transition temperatures and
the  equilibrium values of the pairing gap and  deformation at 
zero temperature.  The linear relationship is
extremely good for  the pairing gap and  quite fair  for the
deformation.  In the  range of nuclei that are studied here, 
it is indeed possible to
estimated very  closely  the value of the pairing transition temperature
$T_\Delta$ for both  neutrons and protons,  as the ground state pairing gap 
can be calculated from the  systematics of binding energy. 
Similarly, it appears that 
the shape transition temperatures can be well estimated 
since the ground state quadrupole deformations 
can be experimentally extracted. The transition 
temperatutres $T_\Delta$
are not too different  from those  calculated 
earlier in a non-relativistic  framework;
the shape transition temperatures $T_c$ however seem to be  higher than the
corresponding values calculated in the (P+Q) model. These higher values of the
shape transition temperatures are  however found to be 
very compatible  with the ones obtained from the 
realistic  Gogny force.

\newpage
\noindent{\bf Figure Captions}
\begin{itemize}
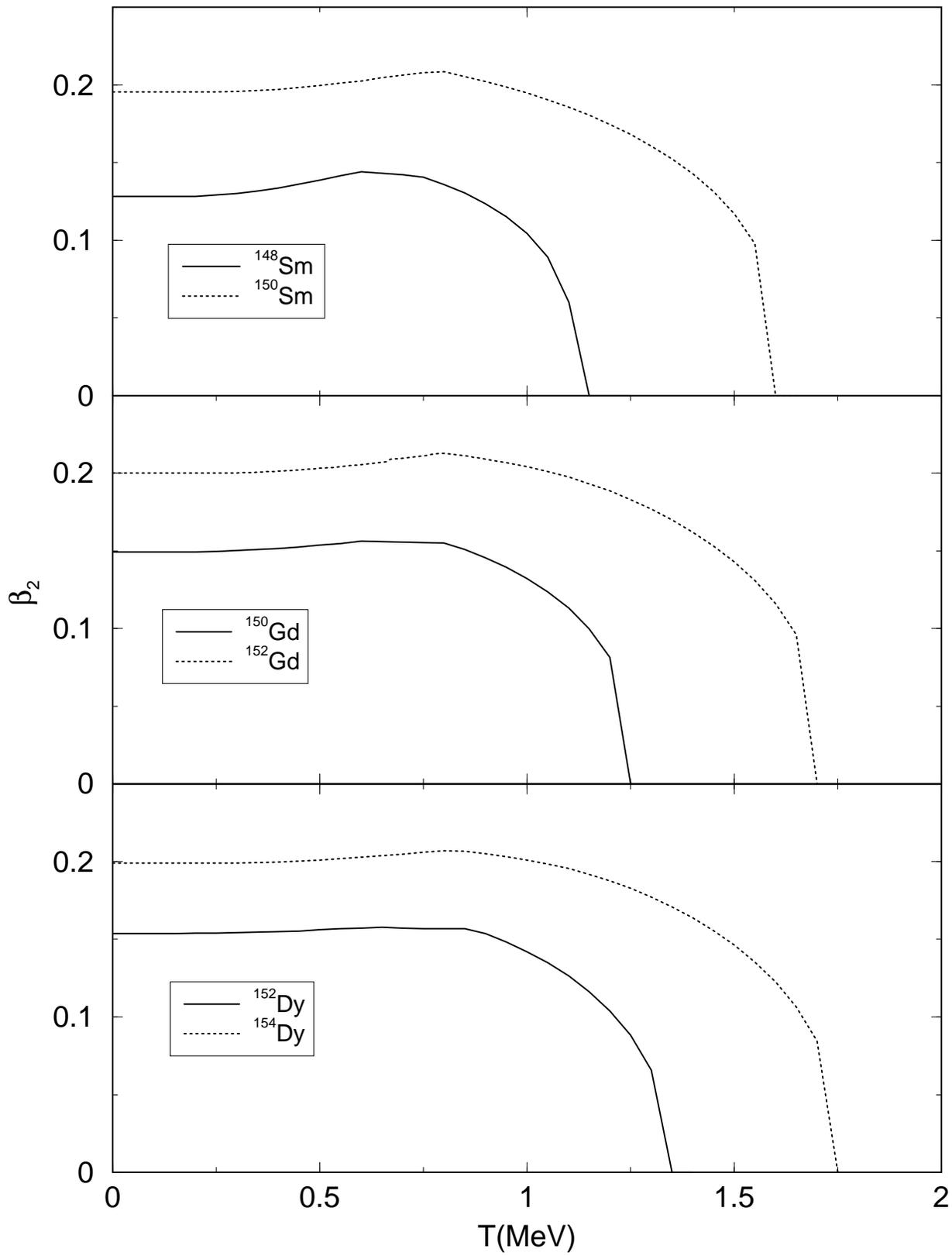
\item[Fig. 1] The evolution of  the quadrupole deformation 
$\beta_2$ as a function
of  temperature for the $Sm$, $Gd$ and $Dy$ isotopes. 
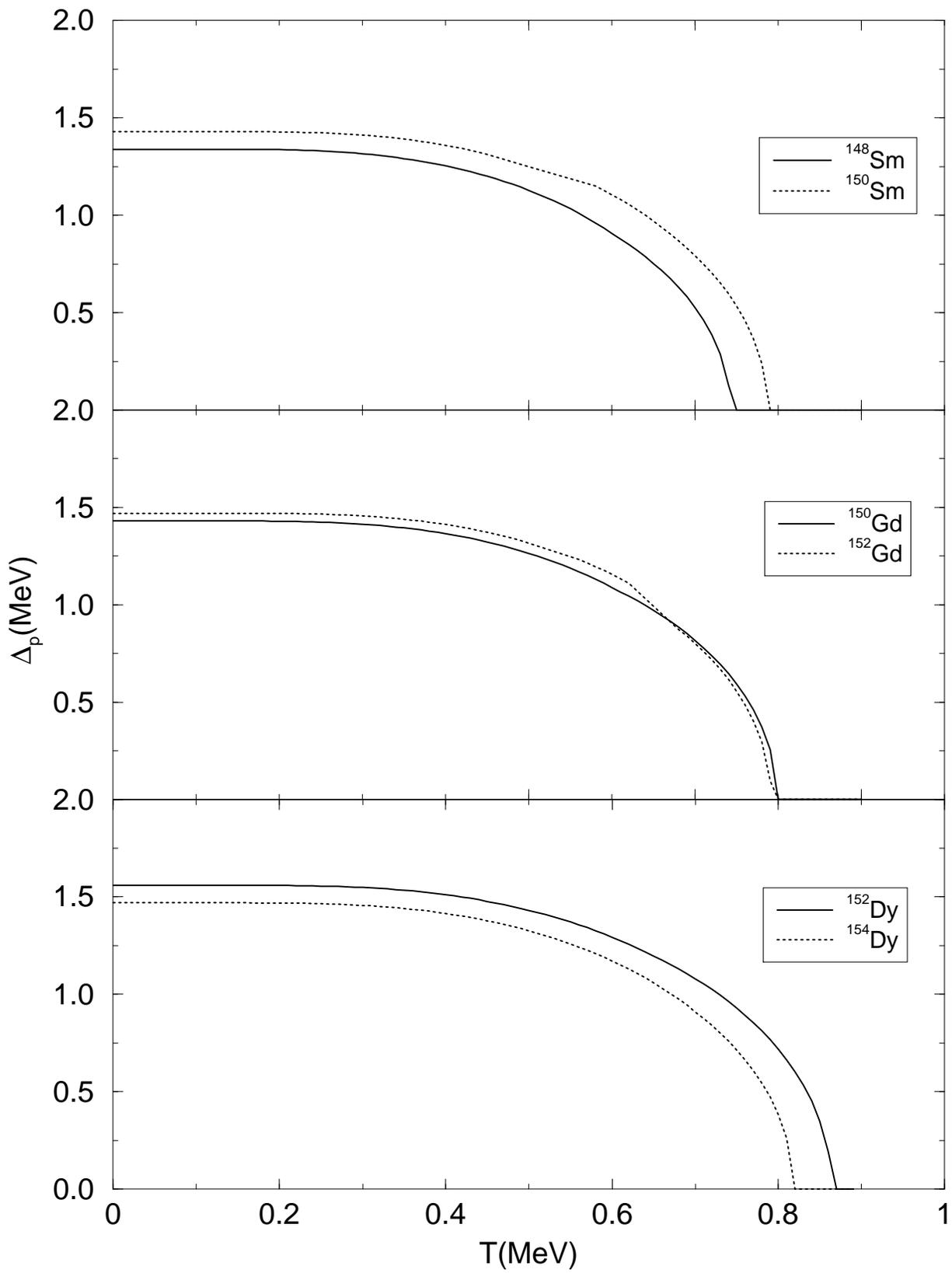
\item[Fig. 2] Temperature evolution of the  proton pairing 
gap for the systems indicated.
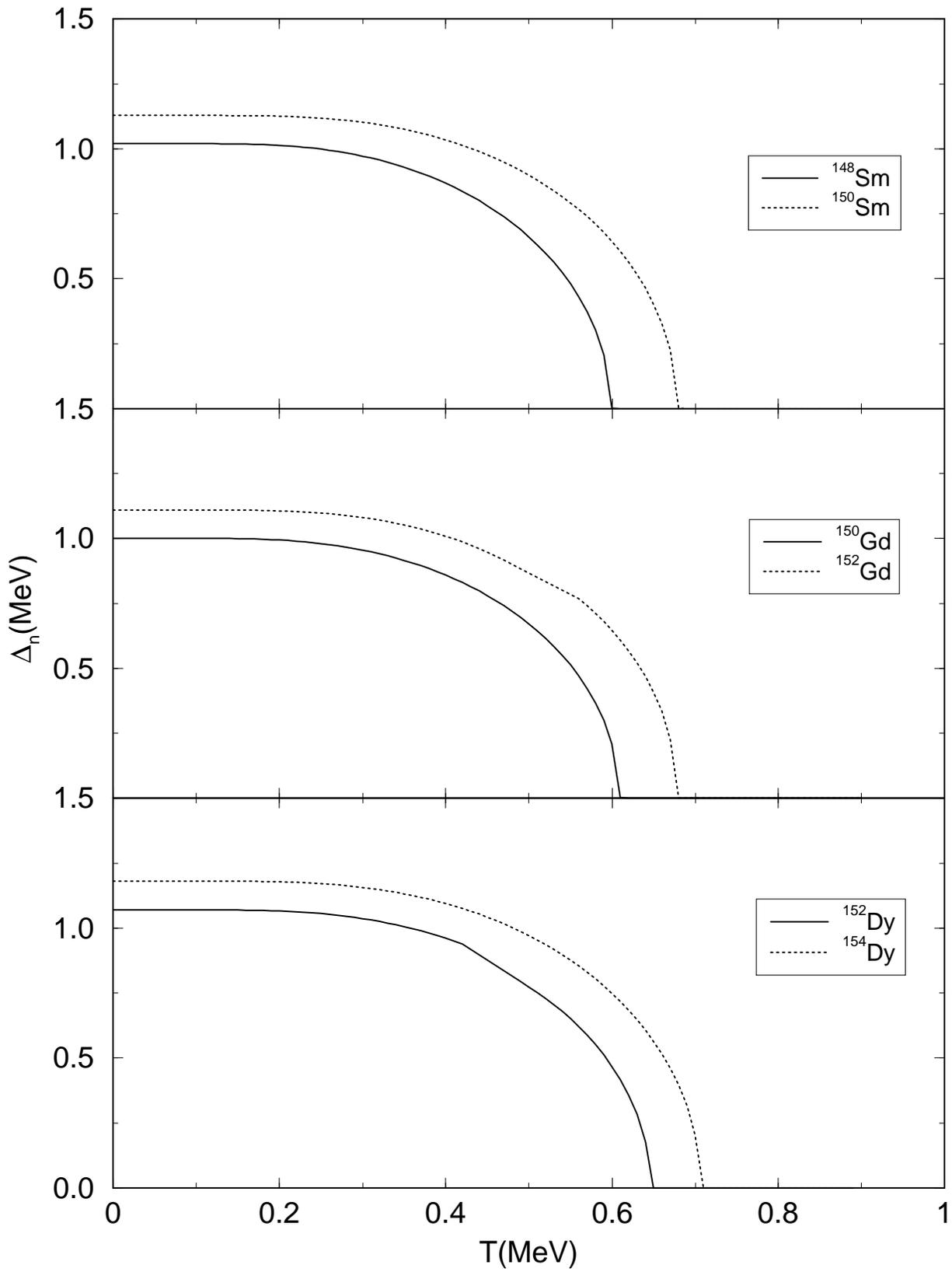
\item[Fig. 3]{Same as in Fig. 2, but for the neutron pairing  gap.}
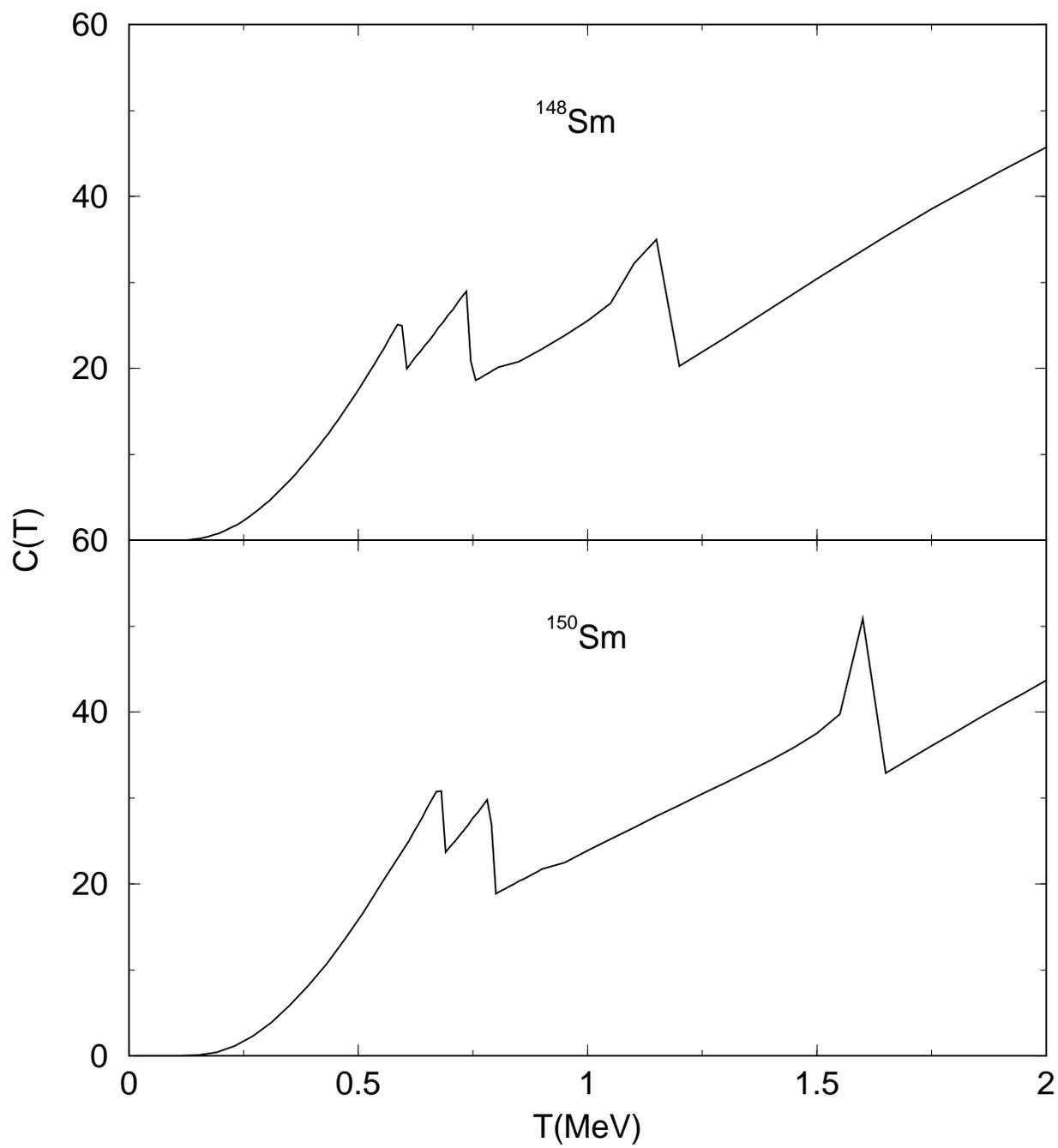
\item[Fig. 4]{Variation of heat capacity  as a function of 
temperature for $^{148}Sm$ and $^{150}Sm$. }
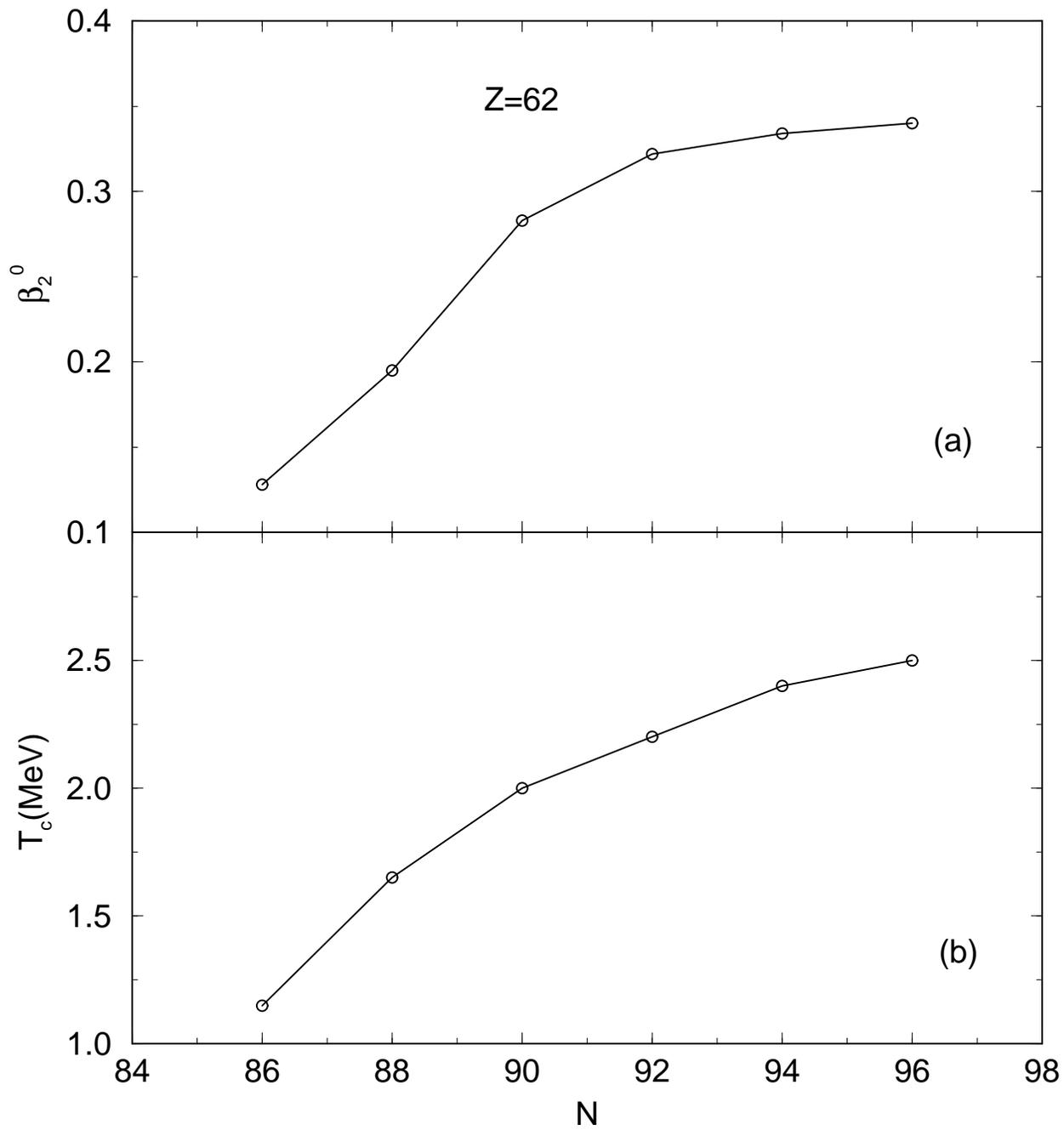
\item[Fig. 5] (a)  Variation of the ground state deformation 
$\beta_2^0 $ with neutron  number $N$ for the
isotopes of  $Sm$ and (b)  the shape transition temperature $T_c$ for different
isotopes of $Sm$.
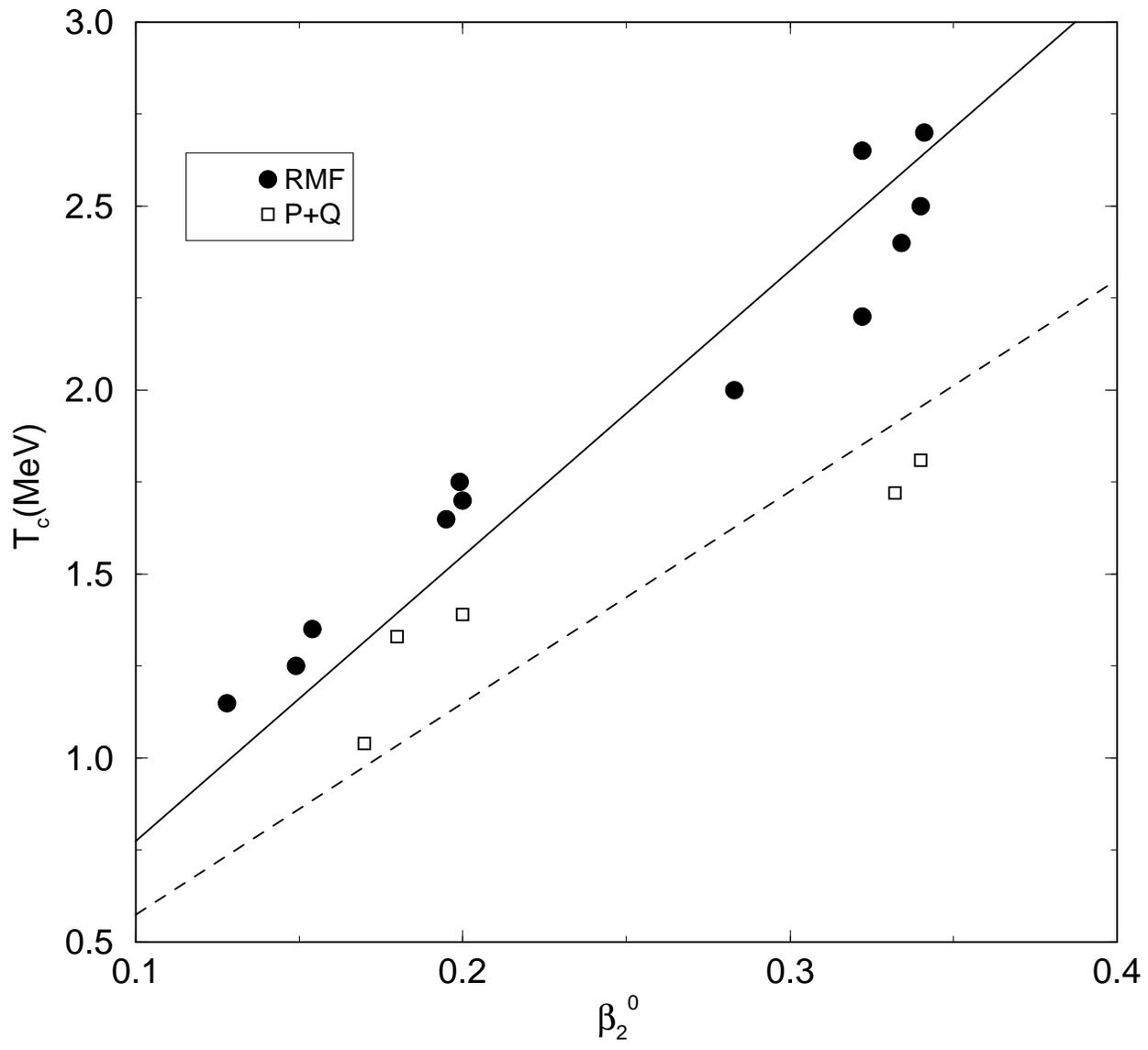
\item[Fig. 6] The shape transition temperature $T_c$ plotted as a 
function of   the ground state
quadrupole deformation $\beta_2^0$ for  different  rare-earth nuclei.
The full circles refer to the results from the RMF theory, the full
line is the linear fit to these. The open squares refer to those  
from the (P+Q) model and the
dashed line is the linear fit  to  these points.
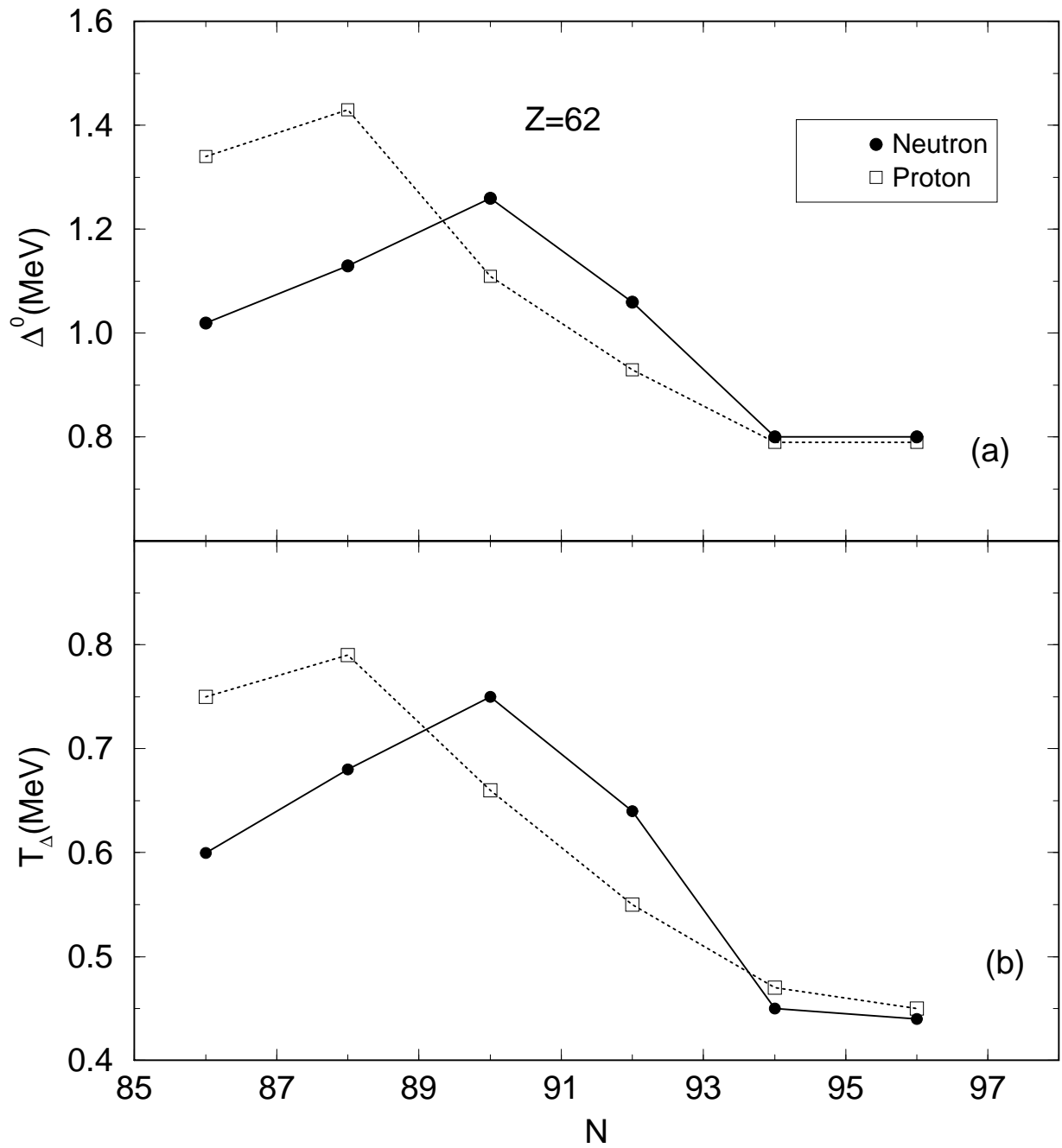
\item[Fig. 7]  (a) The ground state neutron and proton pairing 
gaps as a function of neutron number
for the $Sm$ isotopes and (b)  the neutron and proton 
pairing transition temperatures
$T_\Delta$ for the different $Sm$ isotopes.
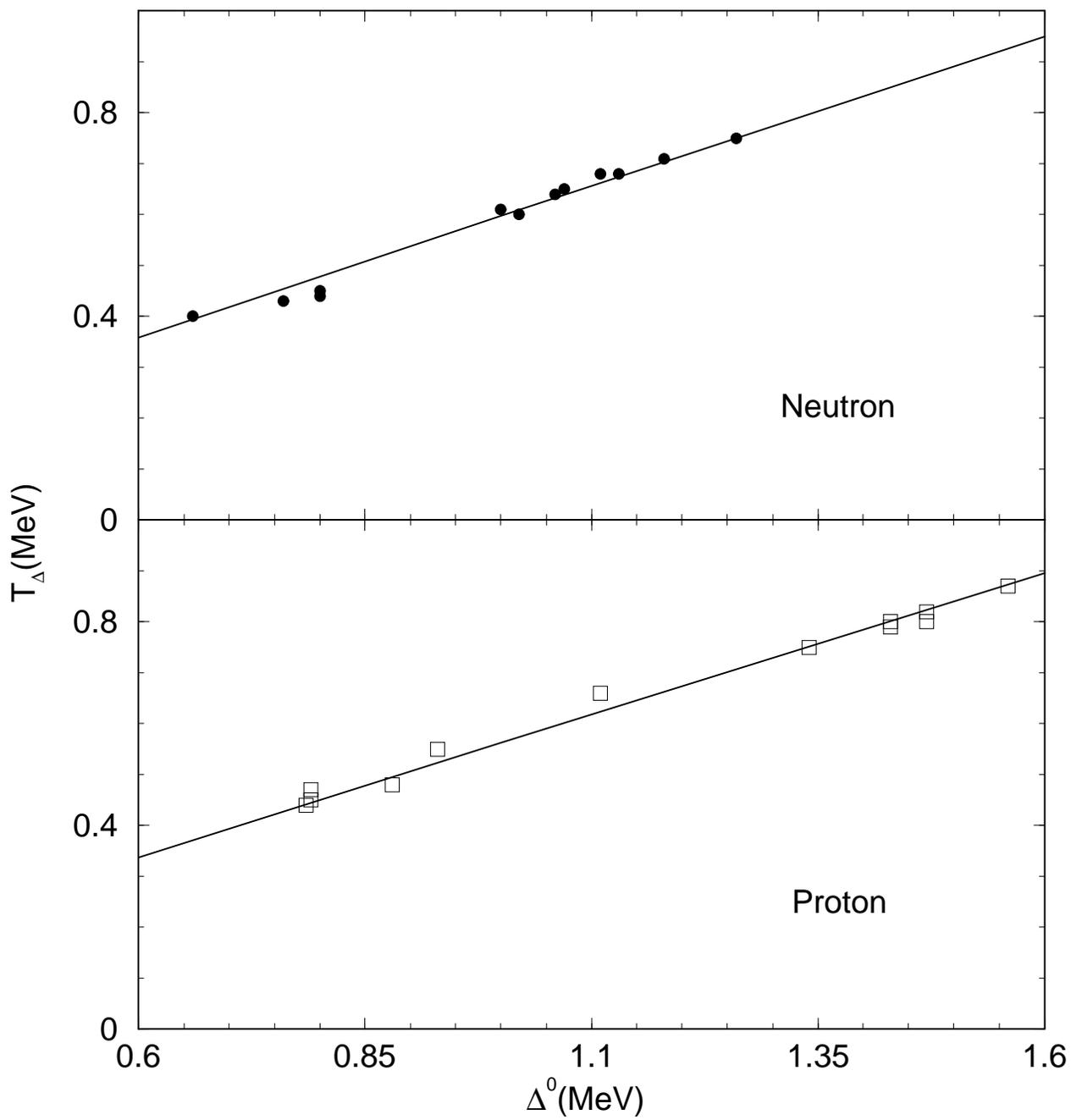
\item[Fig. 8] The correlation between the  transition temperature 
$T_\Delta$ and  the ground state pairing gap $\Delta^0$
for different nuclei in the rare-earth region.
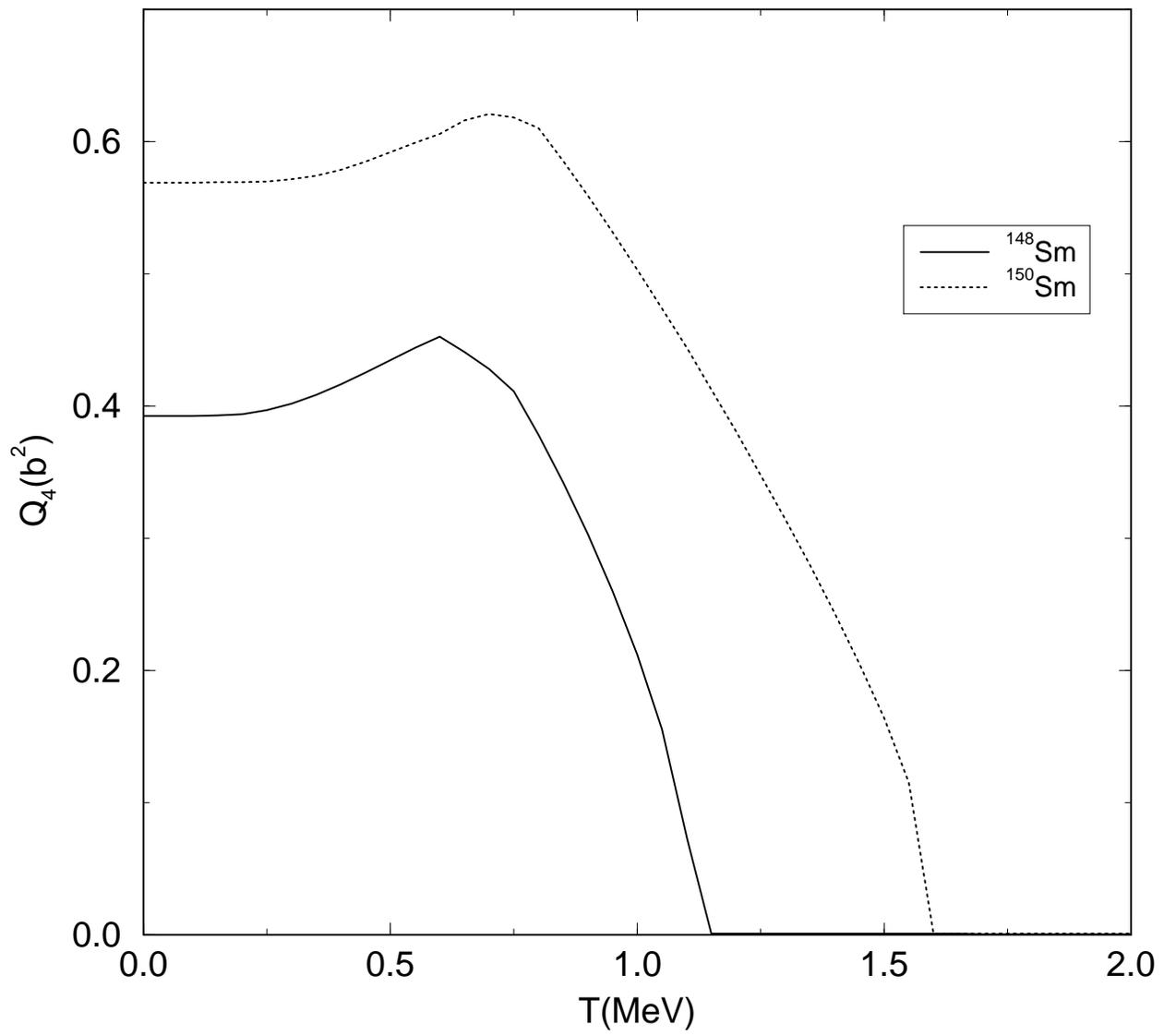
\item[Fig. 9] Temperature evolution of the hexadecapole moment 
(in units of barn$^2$) for the  nuclei 
$^{148}Sm$ and $^{150}Sm$.
\end{itemize}

\end{document}